\documentclass{aa}
\usepackage{graphicx}
\usepackage{amsmath,amssymb}
\usepackage{times}

\begin{document}

\title{Disentangling telluric lines in stellar spectra}
\titlerunning{Disentangling telluric lines}
\author{P. Hadrava}
\authorrunning{P. Hadrava}
\institute{
 Astronomick\'y \'ustav, Akademie v\v{e}d \v{C}esk\'e
 republiky, CZ-251 65 Ond\v{r}ejov, Czech Republic
}
\date{} 

\abstract{
The use of a method of spectra disentangling for telluric lines is
explained in detail, with a particular emphasis on high-precision
radial-velocity measurements for the search for extrasolar planets.
New improvements to the method are introduced.

}
\maketitle

\section{Introduction}\label{Intr}

Ground-based spectroscopy of astrophysical objects is influenced by
telluric lines, which are often pronounced especially in the red and
infrared region. These lines have to be taken into account for numerous
reasons, the most important of which are:
(i) the need to clean the stellar spectra of this additional absorption,
(ii) the possibility to check or to improve the definition of the
wavelength scale in the spectra (see, e.g., Griffin and Griffin 1973,
Greimel and Yang 1999) and
(iii) a study of the Earth's atmosphere (cf. Griffin 2005).
While in the first case the telluric absorption is harmful,
in the other two cases it is useful. For the purpose (ii) not only are
the natural atmospheric lines used but artificially generated
lines from hydrogen-fluoride, iodine or other absorption cells (cf. e.g.
Campbell 1983, Butler et al. 1996) can also be imprinted on the observed
spectra.
Because the problems of the data processing methods discussed here
for absorption cells and atmospheric lines are in principle equivalent,
we shall refer to both cases as the ``telluric lines".

 In the above items it is necessary to accurately separate
the telluric lines from the (`stellar') spectrum of the observed
background object. Several methods have been developed for this
purpose with different advantages and limitations. In principle,
the telluric lines can be distinguished from the stellar lines using
a proper model of their wavelengths, ratios of strengths and line
profiles (cf. e.g. Adelman et al. 1996) or by their variability
in wavelength (due to the annual motion of the Earth) or in strength
(due to variations of the air mass and humidity).

 The method of disentangling spectra of multiple stars (Simon and
Sturm 1994, Hadrava 1995, P1 hereafter) generalized for variable
line-strengths (Hadrava 1997, P2 hereafter) is suitable to separate
telluric lines.
This possibility has been incorporated in the code KOREL for
Fourier disentangling (Hadrava 2004b, P3 hereafter) and has been
widely used (e.g. Harmanec et al. 1997 etc.).

 Hrudkov\'{a} and Harmanec (2005, HH hereafter) applied the KOREL
disentangling of telluric lines to $\alpha$~Boo and advertised
its usefulness for studies of late-type single stars.
Unfortunately, they had not properly taken into account explanations
of the method (Hadrava 2004a or P3) and give some misleading information.
To prevent further confusion we give the explanation more explicitly
here. In addition, some new possibilities and improvements of the method are presented.

\section{Disentangling of telluric lines using KOREL}\label{Velocity}
 The disentangling of spectra is made possible by the
variations of mutual Doppler shifts, and thus works equally for
different components of a multiple stellar system or for
telluric lines moving annually with respect to single or multiple
stars. Telluric, interstellar and binary spectra can be
disentangled simultaneously.
However, because the telluric lines also vary in strength,
their disentangling became possible in practice only after
a generalization enabling one to take into account that line-strength
variation (cf. P2). On the other hand, this variability in  strength
is a helpful feature in disentangling telluric lines.

 In the first applications, the orbital parameters of telluric
lines were converged by the disentangling method. However, because
these parameters are given by the annual motion of the Earth and
by the ecliptic coordinates of the observed object, which are known,
it is advantageous to use the constraints on these parameters
as facilitated by the code PREKOR (cf. P3 for details).
The observed stellar spectra are usually reduced to the heliocentric
wavelength scale. In this case the fictitious centre of mass
of the Earth's relative motion (i.e. orbit No. 3 if the telluric
lines are identified with component No. 5 -- cf. Fig. 1 in P1)
with respect to the star or stellar system is put close to the star,
imposing a very small $K$-velocity of the star, and a proportionally small
value of the mass ratio $q_{3}\equiv m_{\rm Earth}/m_{\rm star}$ is chosen
to give the proper amplitude $K_{\rm Earth}$ of the telluric lines.
The value $K_{\rm star}=1$m/s is preset and the corresponding value of
$q_{3}$ is calculated from the coordinates of the star by the code PREKOR,
but users may rescale both these numbers from PREKOR to keep
$K_{\rm star}$ negligible if a higher accuracy is desirable.
This can be necessary especially in applications that search for extrasolar
planets or other low-mass companions, when the lines of an absorption cell
can also be superimposed on the stellar spectrum to
provide a more precise wavelength scale (as mentioned in Section \ref{Intr}).
However, for such studies requiring high precision, the Keplerian orbit
is an insufficient approximation for the velocity of the Earth and a more
accurate approximation, which takes into account corrections for other
planets in the Solar system, has to be involved. For this purpose
the PREKOR code includes the subroutine kindly provided by C. Ron and
J. Vondr\'{a}k (Ron and Vondr\'{a}k 1986) which calculates the vector
of the Earth's velocity with a precision of 0.02 m/s.
The disentangling can be then performed with the velocities of the telluric
iodine lines unbound to a Keplerian orbit (as described in P3) but fixed
by this more precise model. Some differences between radial velocities
of telluric lines in individual exposures found by KOREL and the
Ron and Vondr\'{a}k model can be caused by instrumental or
data-processing errors in determining the wavelength scale.
The whole spectrum can be then shifted by a new run of PREKOR.
In principle, a version of the disentangling method with self-adjustment
of wavelengths could be developed for cases with well-defined
telluric/iodine lines in all spectra.

 An alternative possibility is to disentangle the spectra
in instrumental, instead of in heliocentric, wavelengths, to set
the $K$-velocity of the whole observed stellar system equal to
the corresponding projection of the Earth's velocity in the ecliptic
on the line of sight, and to formally set the mass ratio
$m_{\rm Earth}/m_{\rm star}$ very large to decrease the $K$-velocity
of telluric lines safely below the resolution capability of the spectra.

\section{The role of line-strength variability}\label{Strength}

\begin{figure}
\setlength{\unitlength}{1mm}
\begin{picture}(88,70)
 \put(10.8,8){\resizebox{57.9mm}{!}{\includegraphics{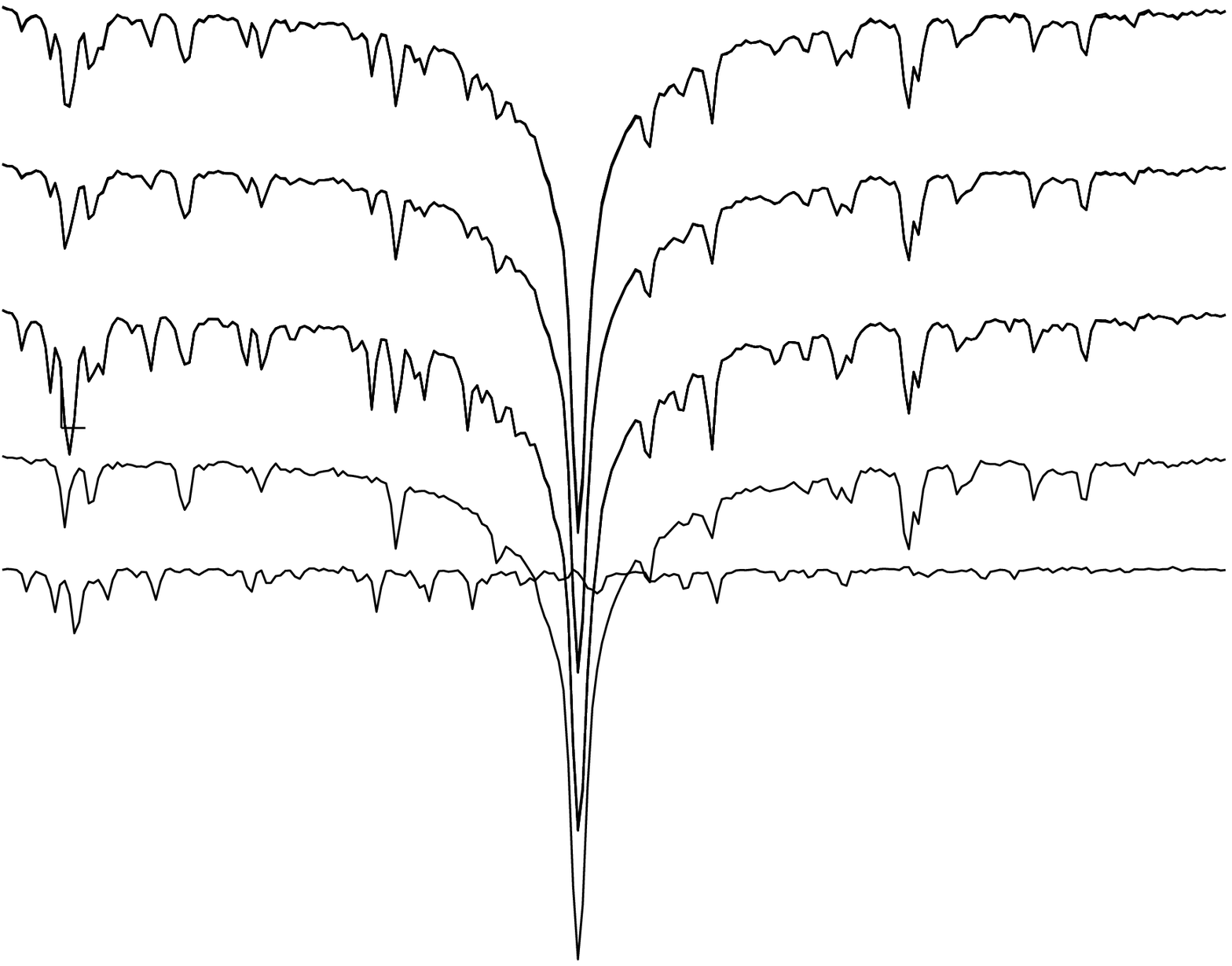}}}
 \put(8,5){\framebox(80,65){}}
 \put(15.218,5){\line(0,1){1.5}}
 \put(22.426,5){\line(0,1){1.5}}
 \put(29.622,5){\line(0,1){1.5}}
 \put(36.807,5){\line(0,1){2.5}}
 \put(32.807,1){655.0}
 \put(43.982,5){\line(0,1){1.5}}
 \put(51.145,5){\line(0,1){1.5}}
 \put(58.298,5){\line(0,1){1.5}}
 \put(65.439,5){\line(0,1){1.5}}
 \put(72.570,5){\line(0,1){2.5}}
 \put(68.570,1){660.0}
 \put(79.690,5){\line(0,1){1.5}}
 \put(86.799,5){\line(0,1){1.5}}
 \put(79,0){$\lambda$\,[nm]}
 \put(0,64.0){$I$}
 \put(8.,66.75){\line(1,0){1.5}}
 \put(8.,63.50){\line(1,0){1.5}}
 \put(8.,60.25){\line(1,0){1.5}}
 \put(8.,57.00){\line(1,0){1.5}}
 \put(8.,53.75){\line(1,0){2.5}}
 \put(1,52.5){1.5}
 \put(8.,50.50){\line(1,0){1.5}}
 \put(8.,47.25){\line(1,0){1.5}}
 \put(8.,44.00){\line(1,0){1.5}}
 \put(8.,40.75){\line(1,0){1.5}}
 \put(8.,37.5){\line(1,0){2.5}}
 \put(1,36.5){1.0}
 \put(8.,34.25){\line(1,0){1.5}}
 \put(8.,31.00){\line(1,0){1.5}}
 \put(8.,27.75){\line(1,0){1.5}}
 \put(8.,24.50){\line(1,0){1.5}}
 \put(8.,21.25){\line(1,0){2.5}}
 \put(1,20.25){0.5}
 \put(8.,18.00){\line(1,0){1.5}}
 \put(8.,14.75){\line(1,0){1.5}}
 \put(8.,11.50){\line(1,0){1.5}}
 \put(8.,8.25){\line(1,0){1.5}}
 \put(88.,66.75){\line(-1,0){1.5}}
 \put(88.,63.50){\line(-1,0){1.5}}
 \put(88.,60.25){\line(-1,0){1.5}}
 \put(88.,57.00){\line(-1,0){1.5}}
 \put(88.,53.75){\line(-1,0){2.5}}
 \put(88.,50.50){\line(-1,0){1.5}}
 \put(88.,47.25){\line(-1,0){1.5}}
 \put(88.,44.00){\line(-1,0){1.5}}
 \put(88.,40.75){\line(-1,0){1.5}}
 \put(88.,37.5){\line(-1,0){2.5}}
 \put(88.,34.25){\line(-1,0){1.5}}
 \put(88.,31.00){\line(-1,0){1.5}}
 \put(88.,27.75){\line(-1,0){1.5}}
 \put(88.,24.50){\line(-1,0){1.5}}
 \put(88.,21.25){\line(-1,0){2.5}}
 \put(88.,18.00){\line(-1,0){1.5}}
 \put(88.,14.75){\line(-1,0){1.5}}
 \put(88.,11.50){\line(-1,0){1.5}}
 \put(88.,8.25){\line(-1,0){1.5}}
\end{picture}
\caption{\label{obr2} The standard KOREL output for three spectra of
 $\tau$~Boo (upper three lines)
 taken on three consecutive nights disentangled as stellar
 (the fourth line) and telluric spectra (bottom line).}
\end{figure}

 It is claimed in HH that ``the disadvantage of the method is that one needs
a number of spectra of the star in question, secured at different months
of the year" to disentangle telluric lines from spectra of a single star.
It is not necessarily the case, as can be demonstrated by the example shown
in Fig.~\ref{obr2}. The three Ond\v{r}ejov RETICON spectra in the
region of H$_{\alpha}$ line of the star $\tau$~Boo, known for 
spectroscopic observations of its planet, were taken within three days.
To exclude the possibility that the disentangling of telluric lines
is due either to the annual change of heliocentric correction of
the radial velocity (which is about 0.6 km/s between the first
and the third exposure) or to the motion of the
star itself (which has a period of about 3 days and is detectable by
disentangling a larger set of spectra), the times of all three
exposures were set equal to the time of the middle one.
Nonetheless, the disentangling is satisfactory. This is owing to the
differences in strengths of the telluric lines, which are obviously
weakest in the second and strongest in the third spectrum.
The difference between two chosen spectra with different strengths
determines the shape of the telluric spectrum $I_{tell}$ (in some
reference scale), which must be then subtracted from each exposure
$I_{obs}(t)$ with a proper multiplicative line-strength factor $s(t)$
to get the stellar spectrum $I_{0}$,
\begin{equation}\label{tel}
 I_{0}=I_{obs}(t)-s(t)I_{tell}\; .
\end{equation}
The solution of factors $s(t)$ is not unique, because $ks(t)$ ($k>0$)
gives the same solution with the rescaled telluric spectrum
$I_{tell}/k$ and $s(t)-\Delta s$ fits the observed spectra equally
by the stellar spectrum $I_{0}+\Delta sI_{tell}$. The strengths $s(t)$
thus cannot be solved directly from equation (7) in P2 or Eq. (30) in P3,
which is singular in this case, but can be converged by a simplex
method which finds some of the existing solutions. By fixing some of the
strengths we can choose the proper solution, or it can be more easily
found by KOREL using even relatively small seasonal differences
in heliocentric radial-velocity corrections (either for stars close to
the poles of the ecliptic or for data covering a short part of the year).
An important condition which is worth satisfying during
the observation is thus obtaining spectra at different air masses.
(Our early experience confirms the seemingly paradoxial fact that
the disentangling of telluric spectra significantly improves
if spectra that are highly blended with telluric lines are included
in the solution.)

 Telluric lines consist of several subsystems whose strength may
vary independently. Typically, lines of O$_2$, N$_2$ and Ar depend mainly
on the air mass, while lines of H$_2$O depend also on the
instantaneous weather (cf. Adelman et al. 1996) and the O$_3$ lines
should be checked for long-term variations (Griffin 2005). Because
these lines have identical radial velocities, they can be disentangled
only by their variable strengths. In KOREL, it is possible by the above
explained method to assign the annual motion of the Earth to both
orbits No. 2 and 3 (and a negligible $K$-velocity to
orbit No. 1), so that all the components in No. 3, 4 and 5 can
correspond to telluric lines (cf. Fig. 1 in P1). However, to separate
the subsystems, good initial estimates of their strengths
are needed, or spectra of at least some of them should be constrained
by an accurate model, as explained below. An alternative method,
introduced by Lynch (2005, cf. Lynch and Polidan 1997) to separate
spectral components that change their line-strengths independently,
could be used for exposures with negligible differences in radial
velocities between the stellar and telluric lines. This also confirms
that the limitation claimed by HH is incorrect. Lynch's method is
based on the SVD/PCA technique and indicates how many independent components
(characterized by their strengths) are present in the data, while this number
must be fixed in the present versions of disentangling. Combination of
both methods can thus be useful for the problems in question.

\section{Non-linearity of telluric lines}\label{Nonlin}
 HH mentioned as a ``principal problem when KOREL is used to disentangle
telluric lines" the question of nonlinearity (Hensberge 2001).
HH then incompletely rephrased the original answer given by the
present author and ignored the recent, more thorough explanations
(Hadrava 2004a or P3).
The `problem' is that the telluric spectrum is not an additive, but
a multiplicative component in the observed spectra, i.e. the observed
spectrum is (cf. Eq.~(31) in P3)
\begin{equation}
 I_{obs}(x,t)=\exp(-\tau(x,t))I_{0}(x,t)\; ,
\end{equation}
where $I_{0}$ is the spectrum of the studied object as seen outside
the Earth's atmosphere and $\tau$ is the optical depth along the
line of sight in the atmosphere. This can be expressed as a sum of $I_{0}$
and a negative telluric component:
\begin{equation}
 I_{tell}=(e^{-\tau}-1)I_{0}\simeq-\tau I_{0}\; ,
\end{equation}
which also depends on $I_{0}$. Consequently, the depth
of the telluric line in Fig.~\ref{obr2} which falls close to the centre
of the stellar H$_{\alpha}$ line is underestimated in the decomposed
spectrum relative to the depths of telluric lines falling on the continuum.
If the disentangled spectra are to be used for spectroscopic studies of
the atmosphere of the Earth, the telluric spectrum must be divided by
the mean stellar spectrum. Such a renormalisation has also to be applied
if we want to use the telluric spectrum disentangled from one star as
a template for constrained disentangling of another star, as will be
explained later. The only discrepancy arises if the telluric line moves
(either due to its own seasonal shifts or due to the orbital motion of
the star) in the part of stellar line-profile with a steep slope,
or if the strength of the line varies, e.g. in an eclipse of the star.
The error in the residuals (O--C) is proportional to the depth of the
telluric line multiplied only by the difference in the local depth of
the stellar line from its mean value, not its whole depth. Possible
deflection of the decomposed spectrum is generally even smaller.

In the simple case of a one-component stellar system (either a single
star or a multiple system dominated by the spectrum of one component) plus
a telluric spectrum the problem of nonlinearity of the telluric component
can be avoided by a logarithmic transformation of the (rectified)
input spectra:
\begin{eqnarray}\nonumber
 I(t,x)\rightarrow I'(t,x)&=&1+{\rm ln}I=\\
 &=&1+{\rm ln}I_{0}(t,x)-\tau(t,x)\; , \label{log}
\end{eqnarray}
where $I_{0}$ is the stellar spectrum and $\tau$ is the
monochromatic optical depth of the Earth's atmosphere (cf. Eq.~(31)
in P3). The corresponding backward transformation should be applied
to the disentangled spectra. This would not be appropriate to
disentangling standard binaries (for which the spectra are
additive, $I_{0}=I_{1}+I_{2}+...$). However, in searching for a
faint component $I_{2}$ in the presence of strong telluric lines
in the spectrum $I_{1}$ of the dominant component, it can
be advantageous to use the linearization ${\rm ln}I_{0}=
{\rm ln}I_{1}+(I_{2}+...)/I_{1}$ and to disentangle the spectra
on a logarithmic scale of intensities. Naturaly, the normalization
of disentangled telluric spectrum described in the previous
paragraph must not be performed if the spectra are transformed
into a logaritmic scale of intensity. Instead, the inverse transforms
of both the telluric and the stellar spectra have to be applied.

\section{Solution instabilities and use of constraints}
 Despite the solutions described above, some problems may appear
or increase when telluric lines are included in the disentangling.
One example is the problem of waves in the disentangled spectra $I_{j}(x)$, due to
instabilities of the lowest Fourier modes $\tilde{I}_{j}(y)$, for which
the corresponding equations (Eqs. (5) in P1, (6) in P2, (20) or (28) in P3)
of general form
\begin{eqnarray}\nonumber
 &&\sum_{j=1}^{n}\left[\sum_{l=1}^{N}w_{l}
  \tilde{\Delta}_{j}(y,t_{l},p)\tilde{\Delta}_{m}^{\ast}(y,t_{l},p)\right]
  \tilde{I}_{j}(y)=\\
 &&\hspace*{25mm}=\sum_{l=1}^{N}w_{l}\tilde{I}(y,t_{l})\tilde{\Delta}_{m}^{\ast}(y,t_{l},p)
\end{eqnarray}
are nearly singular (here $I$ are the spectra observed at time $t_{l}$,
$\Delta_{j}$ are broadening functions for component $j$ -- cf. P1, P2, P3
for details).
These instabilities often increase with a new component such as a
telluric spectrum, at least until the appropriate strengths
or orbital parameters are well approximated. Possible errors in
rectification or real non-Doppler variations of stellar spectra
may force the solution to use its new freedom to fit these
features instead of the true telluric spectrum. Two hints are
recommended in this case: (i) determine the strengths of
telluric lines from a spectral region where they are dominant
(e.g. the short-wavelength region shown in Fig.~\ref{obr2})
and fix them in other regions at least until the
other parameters are not iterated further, or (ii) use
the fact that the telluric lines are mostly narrow, i.e.
contained in higher Fourier modes, and filter out the lowest
modes at least until the correct solution is approached in higher modes.

 Quite often some problems caused by the rectification, line-profile
variability or an unresolved component of the spectrum persist in
centres of deep stellar lines and a small contribution
to those lines then appears in the telluric spectrum, which is
otherwise reasonable (with the correct lines on a flat continuum) in
the far wings of the stellar lines and in the stellar continuum. In
such a case, use of a recently developed `constrained disentangling'
procedure can be helpful. This will be described elsewhere (Hadrava 2005).
For our present purpose it is necessary to know that the observed spectra
can be approximated by the form
\begin{eqnarray}\nonumber
 I(x,t)&=&\sum_{j=1}^{m}I_{j}(x)\ast\Delta_{j}(x,t,p)+\\
 &&+  \sum_{j=m+1}^{n}J_{j}(x)\ast\Delta_{j}(x,t,p)\; ,
\end{eqnarray}
where the first sum includes the component spectra $I_{j}$ which are unknown
and have to be isolated by the disentangling and the second one includes
those spectra which are known and constrained by some templates $J_{j}$
(all component spectra being Doppler shifted or broadened by broadening
functions $\Delta_{j}$, the parameters $p$ of which may be also solved
by disentangling).
The shape of the telluric spectrum can thus be constrained in this sense,
either removing the residuals of stellar lines from the preliminarily
disentangled spectra by their additional rectification, or
directly by using a template for the telluric spectrum obtained
from solutions for other stars (with the abovementioned
renormalization). With these constrained telluric spectra, the
solution may be forced to divide the previous artifacts between
the different disentangled stellar spectra (and to give better mean
line-profiles), or to put the discrepancies into residual spectra,
from where their nature can be judged.

\vspace*{3mm}

{\it Acknowledgements.} The author would like to thank the referee,
Elizabeth Griffin, for helpful comments.
The inspiring discussions with D. Holmgren, J. Kub\'{a}t and D. Lynch
are also acknowledged. Thanks are due to colleagues C. Ron and
J. Vondr\'{a}k for providing their procedure for calculation of
the Earth's velocity.  The observations of $\tau$~Boo by
H. Bo\v{z}i\'{c} are also highly appreciated. This work has been
supported by project AV0Z10030501.

\label{eDruhy}

\vfill

\end{document}